\documentclass[final,5pt,times,review]{elsarticle}

\usepackage{amssymb}
\usepackage{xurl}

\usepackage{lineno}

\journal{Nuclear Instruments and Methods A}

\begin{document}

\begin{frontmatter}


\title{Pushing the high count rate limits of scintillation detectors for challenging neutron-capture experiments}


\author[1]{ J.~Balibrea Correa } %
\ead{javier.balibrea@ific.uv.es}
\cortext[cor1]{Corresponding author}
\author[1]{ J.~Lerendegui-Marco } %
\author[1]{ V.~Babiano-Suarez } %
\author[1]{ C.~Domingo-Pardo } %
\author[1]{ I.~Ladarescu } %
\author[1]{ A.~Tarife\~{n}o-Saldivia } %
\author[1]{ G. de la Fuente-Rosales } %
\author[3]{ V.~Alcayne } %
\author[3]{ D.~Cano-Ott } %
\author[3]{ E.~Gonz\'{a}lez-Romero } %
\author[3]{ T.~Mart\'{\i}nez } %
% ...
%


\address[1]{Instituto de F\'{\i}sica Corpuscular, CSIC - Universidad de Valencia, Spain}
\address[3]{Centro de Investigaciones Energ\'{e}ticas Medioambientales y Tecnol\'{o}gicas (CIEMAT), Spain} %
% ...
%

\begin{abstract}
%
One of the critical aspects for the accurate determination of neutron capture cross sections when combining time-of-flight and total energy detector techniques is the characterization and control of systematic uncertainties associated to the measuring devices. In this work we explore the most conspicuous effects associated to harsh count rate conditions: dead-time and pile-up effects. Both effects, when not properly treated, can lead  to large systematic uncertainties and bias in the determination of neutron cross sections. In the majority of neutron capture measurements carried out at the CERN n\_TOF facility, the detectors of choice are the C$_{6}$D$_{6}$ liquid-based either in form of large-volume cells or recently commissioned sTED detector array, consisting of much smaller-volume modules. To account for the aforementioned effects, we introduce a Monte Carlo model for these detectors mimicking harsh count rate conditions similar to those happening at the CERN n\_TOF 20~m fligth path vertical measuring station. The model parameters are extracted by comparison with the experimental data taken at the same facility during 2022 experimental campaign. We propose a novel methodology to consider both, dead-time and pile-up effects simultaneously for these fast detectors and check the applicability to experimental data from $^{197}$Au($n$,$\gamma$), including the saturated 4.9~eV resonance which is an important component of normalization for neutron cross section measurements.
\end{abstract}




\end{frontmatter}

\section{Introduction}

The CERN neutron Time-of-Flight (ToF) facility (n\_TOF) is devoted to measurements of neutron-induced cross sections of interest for {\it s}- and {\it r}- stellar processes and Big-Bang nucleosynthesis studies, nuclear technology and medical applications~\cite{Chiaveri2016,Mengoni2019}. The neutron beam at n\_TOF is produced by means of spallation reactions made from 20~GeV/c, low intensity (LI) ($\sim$3$\times$10$^{12}$ particle) and high intensity (HI) ($\sim$8$\times$10$^{12}$particle) proton pulses delivered by the CERN Proton Synchrotron to a lead spallation target~\cite{Esposito2021}. During the spallation process, high energy neutrons are produced and moderated into a white neutron energy spectrum covering from thermal energies ($\sim$0.025~eV) up to hundreds of MeV. Neutrons reach two Experimental AReas located 185~m (EAR1)~\cite{Guerrero2013} and 20~m (EAR2)~\cite{Weiss2015} from the spallation target where the detection devices are placed. A third experimental area called NEAR has been recently installed at only 3~m from the spallation target and it is intended to be used for activation measurements and radiation-damage studies~\cite{Gervino2022,Patronis2022,Cechetto22,Ferrari22}.

The recent upgrade of the spallation source~\cite{Esposito2021} has significantly improved the quality of the neutron beam for ($n,\gamma$) experiments at EAR2~\cite{Lerendegui2023}. In particular, the 50\% enhancement in instantaneous neutron flux and the improved resolution function have been found especially relevant for the measurement of small mass (few mg) and highly radioactive samples (several MBq). Two recent examples are the measurements of $^{79}$Se($n$,$\gamma$) and $^{94}$Nb($n$,$\gamma$) utilizing samples of only $2\times10^{19}$ and $2.24\times10^{18}$ atoms with overall $\gamma$-emitter activities of 7 MBq and 10.1 MBq, respectively~\cite{Lerendegui2023,cdsNb94,cdsSe79,Balibrea22-NPA,Domingo-Pardo2023}. However, full exploitation of high-quality neutron beams requires parallel efforts in developing instrumentation and methods that can cope with the high count rate conditions encountered in such ($n,\gamma$) experiments at EAR2.

At CERN n\_TOF a significant amount of neutron capture cross section measurements are made using C$_{6}$D$_{6}$-based liquid scintillators. When combined with the Pulse-Height Weighting Technique (PHWT)~\cite{MacKlin1967,Abbondanno2004,Borella07}, for convenience, these detectors are usually referred to as Total-Energy Detectors (TEDs) because they mimic the performance of an ideal TED Moxon-Rae detector~\cite{MoxonRae}. Their fast time response (3~ns risetime), narrow pulse-width (10-20~ns FWHM) and low neutron sensitivity ($10^{-4}$)~\cite{PLAG2003} are the key features for their systematic use in this kind of measurements. However, most  C$_{6}$D$_{6}$ detectors utilized over the last 20 years were based on relatively large cell volumes, of about 600~mL~\cite{Abbondanno2004,Borella07}, which were later enlarged to 1~L for optimization of beam-time and counting statistics~\cite{PLAG2003,Mastinu2013}. At the very high instantaneous neutron flux of EAR2 such large detection volumes lead to prohibitive pile-up conditions and dead-time effects. Both effects are critical in the accurate determination of ($n$,$\gamma$) cross sections using TED detectors, especially when aiming at a few percent systematic uncertainty in the final cross section~\cite{Abbondanno2004}. On the one hand, dead-time reduces the number of detected events respect to their actual number and consequently a systematic underestimation of the experimental cross section if the effect is not properly treated. On the other hand, a non identified or non-resolved pile-up event may lead to a wrong deposited energy assignment, which in turn has a direct impact when the PHWT is applied~\cite{MacKlin1967,Abbondanno2004,Borella07}.

One option to mitigate these experimental effects would be to place the detectors sufficiently far from the capture sample. However, this approach reduces also the signal-to-background ratio as the background is virtually constant in the EAR2 and severely limits the attainable detection sensitivity. Alternatively, smaller detection volumes working together as an array can be used, so that the overall count rate is shared among more (smaller) detection units and thus reduce the intrinsic instantaneous count rate per detector. The latter approach has been implemented in the so-called segmented Total Energy Detector (sTED)~\cite{Alcayne23}, which have been successfully used as array of several modules in the two aforementioned capture experiments~\cite{Lerendegui2023,cdsNb94,cdsSe79,Balibrea22-NPA,Domingo-Pardo2023}. Individual sTED modules have a sensitive volume of only 49~ml, which is about a factor of 10-20 smaller than large-volume C$_{6}$D$_{6}$ detectors. However, their combination in a compact array around the sample, allows one to significantly increase its signal-to-background ratio while achieving a similar total efficiency as conventional large-volume C$_6$D$_6$ detectors. For the sake of clarity, the latter will be labelled in this work as C$_{6}$D$_{6}$ whereas sTED modules will be explicitly named.

For ($n$,$\gamma$) cross section measurements in the resolved-resonance region, which is characterized by strongly varying count rates as a function of the measured time of flight, the aforementioned dead-time and pulse pile-up effects represent a conspicuous source of systematic error. At n\_TOF EAR1, the paralyzable dead-time model~\cite{Knoll1979} has been found sufficiently reliable for C$_{6}$D$_{6}$ detectors~\cite{Lerendegui2018,Gawlik2021,Dietz2021}. However, even for small-volume sTED modules, the count rate at EAR2 can easily be significantly higher than reached in EAR1 and requires a specific methodology for experiments, where high count rates are expected, such as experiments with short sample-detector distance or when targets cover a large fraction of the neutron beam. This work describes such methodology, which is based on a novel combination of Monte Carlo (MC)-based dead-time and pile-up generators for TED detectors. The proposed approach provides a simultaneous treatment of both effects in a consistent fashion, and it enables one to handle rather large count rate conditions at least up to 6 and 8 MHz for C$_{6}$D$_{6}$ and sTEDs, respectively. This methodology is demonstrated by applying it to both C$_{6}$D$_{6}$ and small-volume sTED modules. The model exploits some of the features readily developed for the n\_TOF Total Absorption Calorimeter~\cite{MENDOZA2014,GUERRERO2015} and adds some additional features to adapt it to the much faster signals of C$_{6}$D$_{6}$-based detectors.

The article is organized as follows. The acquisition system and data processing pipeline at the CERN n\_TOF facility are briefly introduced in Sec.~\ref{sec:Pre}. Sec.~\ref{sec:Model} presents the new MC dead-time and pile-up model for a fixed count rate situation. The detection probability of two consecutive signals and how to model it from the measured experimental data is discussed in Sec.~\ref{sec:dead-time}, while Sec.~\ref{sec:pile-up_model} describes in detail the pile-up model itself. Sec.~\ref{sec:correction} elaborates on the proposed correction to account for both dead-time and pile-up effects and their actual limitations in terms of count rate. In Sec.~\ref{sec:corr_variation} is described the methodology for a count rate variation scenario. Sec.~\ref{sec:results} illustrates the performance of the new model in the case of the $^{197}$Au($n$,$\gamma$) reaction measured at n\_TOF EAR2 using both C$_{6}$D$_{6}$-based detection systems. Finally, Sec.~\ref{sec:Summary_and_Conclusions} summarizes the main conclusions of this work.  

\section{Acquisition and data processing at n\_TOF}\label{sec:Pre}

At CERN n\_TOF the signals from the different detectors are digitized using {\sc{Teledyne}} SP devices (model ADQ14DC-4C-VG) of four channels per module, with 2 GB memory per channel, 500 (700)~MHz bandwidth at -1 (-3)~dB, and 14~bits flash ADCs. In terms of data-acquisition itself, this digital acquisition approach represents a virtually dead-time free solution~\cite{teledyne2023}. The data buffers are stored using the CERN Advanced STORage manager (CASTOR)~\cite{CASTOR} and analyzed off-line by applying a well established multipurpose Pulse-Shape Analysis (PSA) routine~\cite{ZUGEC2016,GUERRERO2008,LIDDICK2012}.

\begin{figure*}
    \centering
    \begin{tabular}{c}
    \includegraphics[width=\columnwidth]{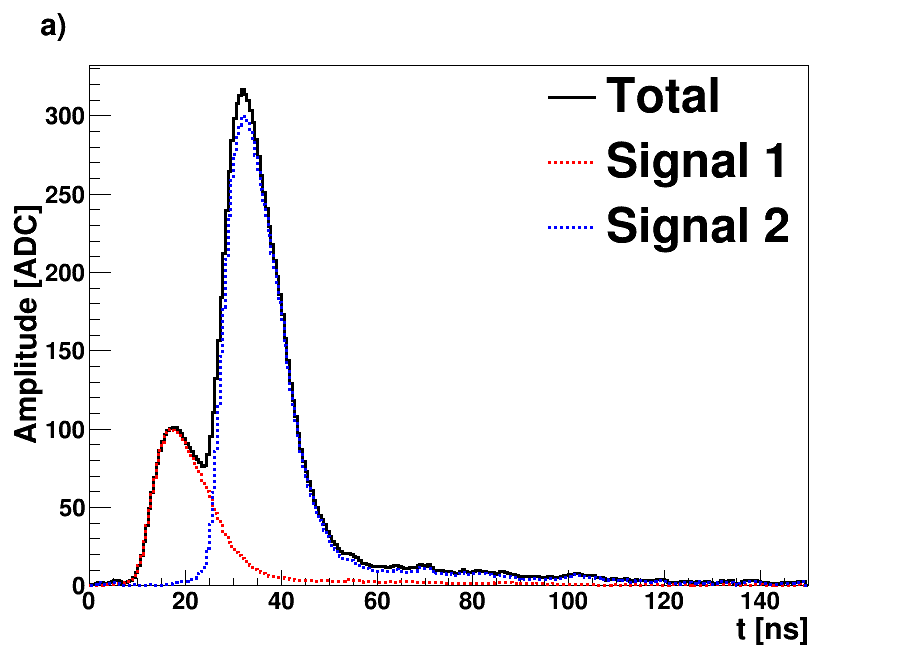} \\
    \includegraphics[width=\columnwidth]{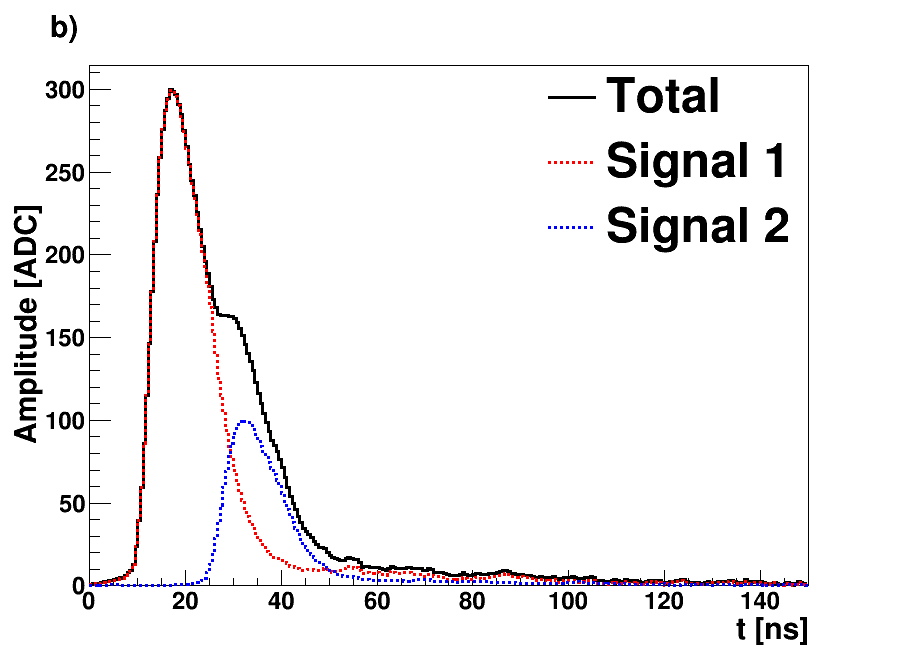} 
    \end{tabular}
    \caption{Illustration of asymmetric C$_{6}$D$_{6}$ detection probability for two signals separated in time by 15~ns. In Panel a) the detection probability for the second signal is 75\%. In panel b) the detection probability for the second signal decrease down to 10\%.}
    \label{fig:Signal}
\end{figure*}

Detector signals are recognized in the data buffers by using a slow derivative algorithm according to the pulse width~\cite{ZUGEC2016}. After the pulse is recognised, the baseline is subtracted and its amplitude, area and time of arrival are determined. The asymmetric pulse shape of liquid scintillation detectors~\cite{MARRONE2002} may introduce limitations in the pulse recognition algorithm, thus producing distortions in the reconstruction of the signal properties for count rates significantly smaller than the inverse value of the detector pulse width. Furthermore, these distortions typically depend on the absolute and relative amplitudes of the two involved events~\cite{MENDOZA2014,GUERRERO2015}. This statement is illustrated in Fig.~\ref{fig:Signal}, which shows two representative situations involving two consecutive signals from C$_{6}$D$_{6}$ with different amplitudes separated in time by 15~ns. 
In both situations the first signal is detected by the PSA routine while, because of the short time difference, the detection probability for the second signal is smaller than 1. For the situation in panel a), the larger amplitude of the second signal leads to a large detection probability (75\%) due to the abrupt change in the slope of the combined signal. However, in panel b) the detection probability of the second signal decreases to almost a negligible value (10\%) since in this situation the slope of the combined signal almost does not change. Such cases introduce reconstruction distortions that are mainly reflected in two different effects: $i)$ dead-time ascribed to the fact that the second signal is not detected by the PSA routine and thus the second event is not accounted for and $ii)$ pile-up corresponding to the fact that the second signal is not identified by the PSA routine but affects to the parameters of the first one by enlarging its apparent area or amplitude. 
In the case of really high count rate, the events registered by the detector may overlap even in the rising part of the signal. In such case, it is impossible by any pulse shape routine to identify the pile-up situation.
These two effects, although negligible in experiments with moderate peak count rates similar to those reached at EAR1, may represent indeed a severe limitation for high count rate situations, like those encountered at EAR2. 

\section{Dead-time, pile-up models and proposed correction}\label{sec:Model}

\subsection{Dead-time and pile-up models in a fixed count rate situation}

The MC model presented here allows to assess dead-time and pile-up effects separately, attending to the different response function of sTED and C$_{6}$D$_{6}$ detectors. The model is inspired by previous works~\cite{MENDOZA2014,GUERRERO2015} related to the BaF$_{2}$ n\_TOF electromagnetic calorimeter~\cite{GUERRERO2009}. The proposed algorithm assumes that the overall count rate is dominated by a single emission source, namely, the $\gamma$-ray cascade following a neutron-capture reaction. In case of additional contributions to overall counting rate, a different approach must be followed, as described in~\cite{MENDOZA2014}. The ingredients of the MC model, in the sequential order of application, are described in the following:

\begin{enumerate}
\item {\it{($n$,$\gamma$) cascade generator:}} The starting points are realistic ($n$,$\gamma$) de-excitation $\gamma$-ray cascades, which mimic the underlying process in a neutron capture experiment. For illustration purposes we use here $^{197}$Au($n$,$\gamma$) capture cascades~\cite{Valenta2022}, generated by \textsc{DICEBOX} code~\cite{BECVAR1998}. Other existing models of Au cascades are likely to provide similar results~\cite{BECVAR1998,TAIN2007,MENDOZA2020}. The output from this stage is a series of neutron capture events, each one consisting of a list of individual $\gamma$-ray energies emitted in a decay.

\item {\it{MC simulation of the "undistorted" detector response to ($n$,$\gamma$) events:}} The geometry of the detection setup is implemented in a C++ application based on the \textsc{Geant4} toolkit~\cite{ALLISON:2016}, mimicking the experiment described in previous works~\cite{Lerendegui2023,Balibrea22-NPA}. In the model, the deposited energy resolution of the different detection systems is included. At this stage each detected $\gamma$-ray event is convolved with the experimental response of the corresponding detector. A list of "undistorted" deposited energies, $E$, is thus generated for each particular detector. In addition, a deposited-energy spectrum $h_{MC}(E)$, free of dead-time and pile-up effects, is constructed as shown by the black dashed line in Fig.~\ref{fig:Deposited_spectra}.

Alternatively one could use the detector response obtained experimentally for ($n$,$\gamma$) cascades at low count rate conditions (i.e. with a very thin sample or reduced beam intensity)~\cite{GUERRERO2015}. However, this approach will be affected by the fraction of the spectrum missing below the detection threshold (typically a few hundreds of~keV) that will contribute both to dead-time and pile-up effects. For instance, a pair of $\gamma$-rays of energy slightly below the detection threshold will not contribute to the undistorted spectra. However, if they are detected within a very short time, the combined signal can exceed the detection threshold and one pulse is recorded. Similarly, a sub-threshold signal can distort amplitude and area of an above-threshold signal. 

\item {\it{Time sampling:}} At this stage the count rate $r$ is introduced in the model. Individual events contributing to $h_{MC}(E)$ are sampled in time assuming a- constant count rate $r$ and standard Poisson probability distribution $P(\Delta t|r)$ of time-differences $\Delta t$ between actual consecutive signals~\cite{MENDOZA2014,GUERRERO2015}

\begin{equation}\label{eq:ProbDeltaT}
    P(\Delta t|r)=e^{-r \Delta t}.
\end{equation}

This step essentially introduces the count rate conditions corresponding to the experiment. From this sampling, a list of deposited energies generated in the previous step and absolute detection times is generated for the quantification of pile-up and dead-time effects.

\item {\it{Pile-up generator:}} The pile-up effect is applied to consecutive signals in the event list from the previous step using the method described in Sec.~\ref{sec:pile-up_model}, thereby modifying the energy of the detected events and removing those that cannot be disentangled by the pulse shape recognition algorithm.

\item {\it{Dead-time generator:}} The detection probability for two consecutive signals is then applied as described in Sec.~\ref{sec:dead-time}. It removes events that can not be identified by the PSA routine. A predefined dead-time function, which is experimentally characterized for each particular detection system, is exploited for this purpose.

\begin{figure}
    \centering
    \begin{tabular}{c}
    \includegraphics[width=\columnwidth]{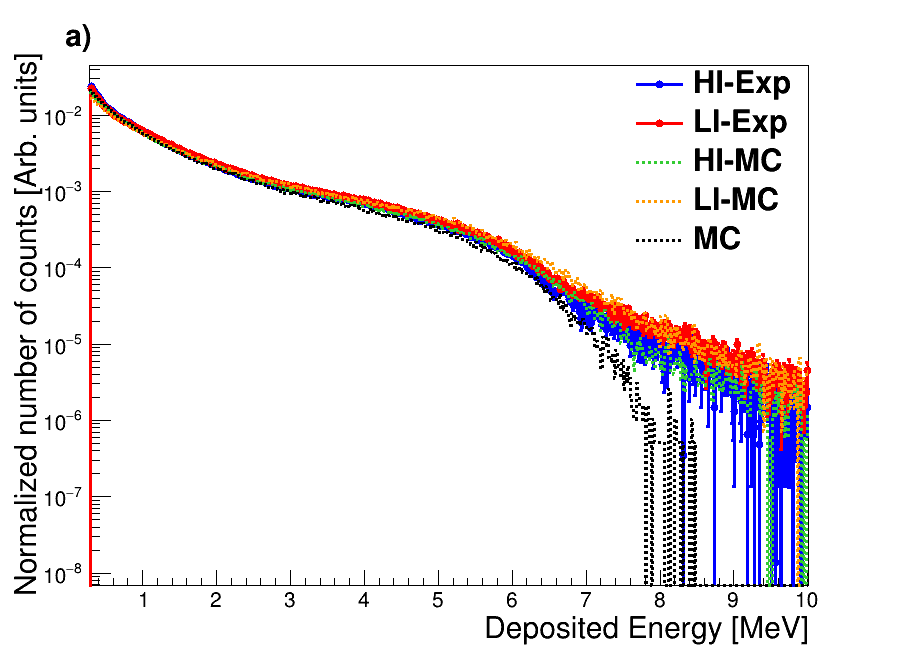} \\
    \includegraphics[width=\columnwidth]{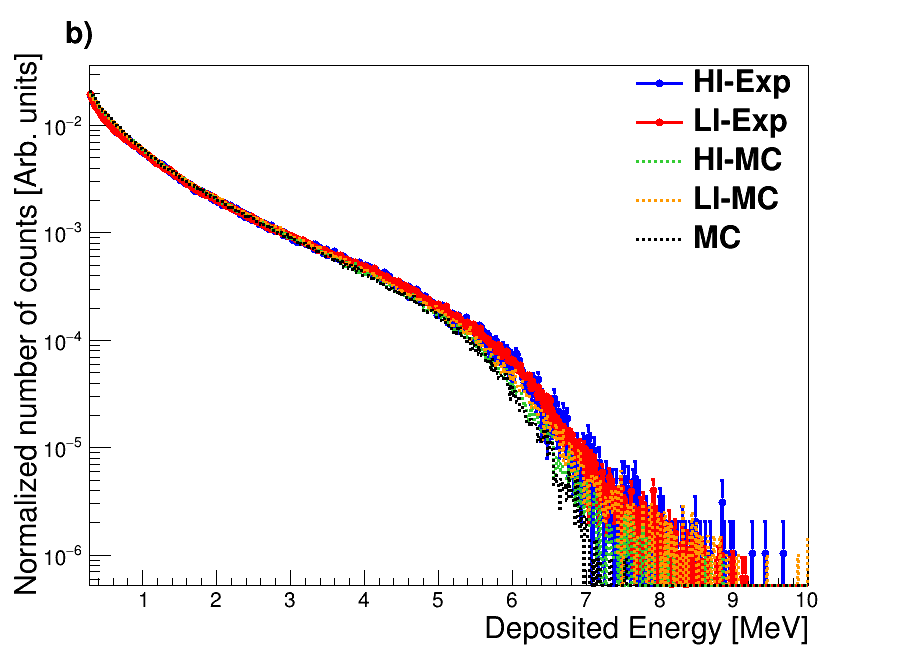} \\
    \end{tabular}
    \caption{Experimental and reconstructed MC deposited energy spectra for C$_{6}$D$_{6}$ (panel a) and sTED (panel b) in the $^{197}$Au($n$,$\gamma$) 4.9~eV saturated resonance during HI (red) and LI (blue) pulses.}
    \label{fig:Deposited_spectra}
\end{figure}

\end{enumerate}

After these steps, the algorithm delivers a realistic list of simulated events, which is used to reconstruct the deposited energy spectrum affected by both dead-time and pile-up effects, $h_{d}(E)$. This spectrum can be directly compared with the registered experimental spectrum. The comparison is made in Fig.~\ref{fig:Deposited_spectra} for C$_{6}$D$_{6}$ (panel a) and sTED (panel b) detectors and separately for LI and HI proton-pulses. The two sets of C$_{6}$D$_{6}$ data shown in Fig.~\ref{fig:Deposited_spectra} correspond to an observed count rate of $\sim$3.2~c/$\mu$s and $\sim$7.5~c/$\mu$s for LI and HI, respectively. In the case of the sTED module, the measured count rates are then $\sim$1.5~c/$\mu$s and $\sim$3.9~c/$\mu$s, respectively. The similar experimental count rate for both detection systems during the measurement, even with the large difference in volume, is due to the distances to the target selected for the experiment. From Fig.~\ref{fig:Deposited_spectra} it becomes clear that if both pile-up and dead-time effects are not properly included in the simulation, the shape of the experimental deposited-energy spectra is not well reproduced by the MC simulation. Thus, both effects become of utmost importance for the correct assessment of the detector efficiency and for determining the capture-yield normalization via the 4.9 eV resonance in $^{197}$Au($n$,$\gamma$)~\cite{MacKlin1969,Massimi2011,SCHILLEBEECKX2012}. 

\subsubsection{Dead-time generator}\label{sec:dead-time}

The C$_{6}$D$_{6}$ and sTED modules fast time response and narrow pulse-width requires the use of a smooth time-difference function for estimating the detection probability of every two consecutive events. A sigmoid function~\cite{bishop2006} has been used in this work as it reasonably describes experimental data. This function depends on the deposited energy (amplitude), $E_{1}$ and $E_{2}$, of the two consecutive signals and their $\Delta t$ as 

\begin{equation}\label{eq:SigmoidDT}
f(\Delta t, E_{1}, E_{2})=\frac{1}{1+e^{-a_{\circ}(E_{1},E_{2})\left(\Delta t-\Delta t_{\circ}(E_{1},E_{2})\right)}}.
\end{equation}

\begin{figure}
    \centering
    \begin{tabular}{c}
    \includegraphics[width=\columnwidth]{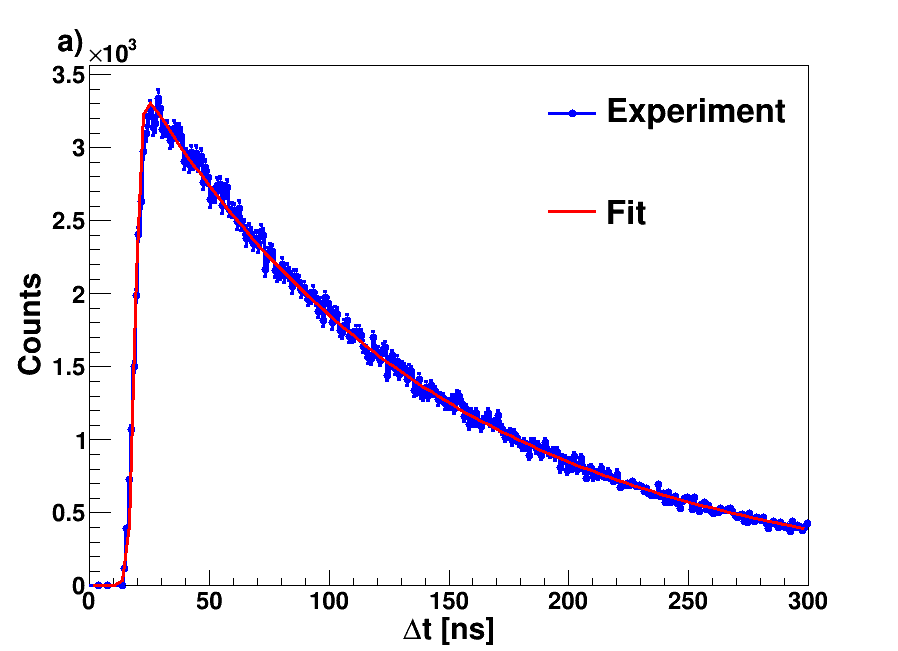} \\
    \includegraphics[width=\columnwidth]{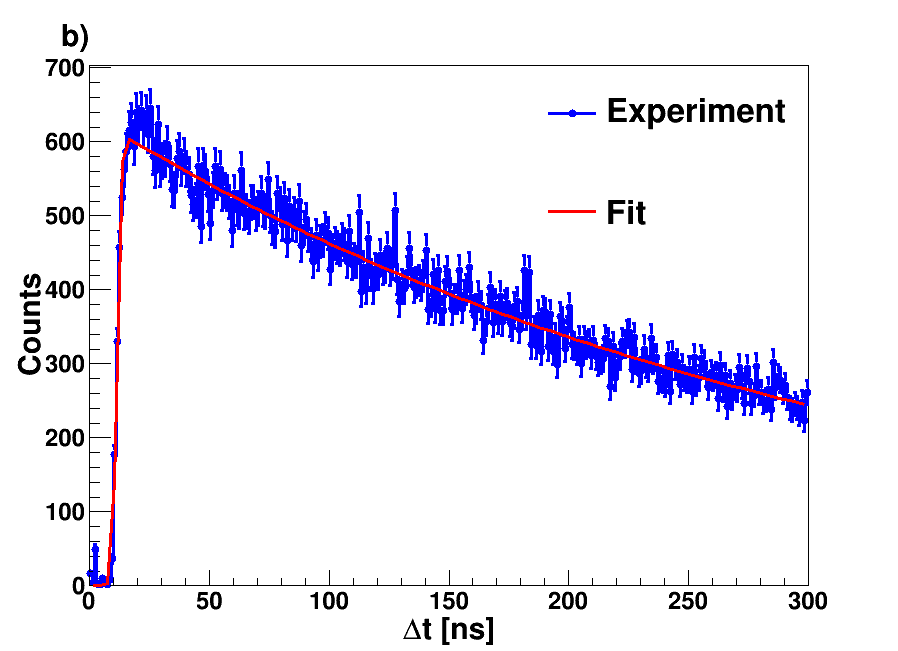} \\
    \end{tabular}
    \caption{Example of $\Delta t$ distribution and analytical fit using deposited energy signals of $E_1$ and $E_2$ between 1 and 2 MeV for C$_{6}$D$_{6}$ detectors (panel a) and sTED module (panel b).}
    \label{fig:DT_sTED}
\end{figure}

The parameters $\Delta t_{\circ}(E_{1},E_{2})$ and $a_{\circ}(E_{1},E_{2})$ define the time difference for a 50\% detection probability of the second signal and the detection-probability rate change for a particular pair of values $E_{1}$ and $E_{2}$. Both $a_\circ$ and $\Delta t_{\circ}$ can be experimentally determined by fitting the measured $\Delta t$ distributions between consecutive signals for any $E_{1}$ and $E_{2}$ to the product of functions given by Eqs.~(\ref{eq:ProbDeltaT}) and (\ref{eq:SigmoidDT}) scaled by a normalization factor $A$

\begin{equation}\label{eq:fitting}
g(\Delta t, E_{1}, E_{2})=\frac{A\times e^{-r_{1}\Delta t}}{1+e^{-a_\circ(E_{1},E_{2})\left(\Delta t-\Delta t_{\circ}(E_{1},E_{2})\right)}}.
\end{equation}
Here, $r_{1}$ is the count rate of the underlying expected exponential time difference distribution associated to a Poisson random process. As indicated by Eq.~(\ref{eq:fitting}), $A$ and $r_{1}$ have no impact on values of $a_{\circ}$ and $\Delta t_{\circ}$ and depend only on ranges of $E_{1}$ and $E_{2}$ used in the fitting procedure. Fig.~\ref{fig:DT_sTED} shows results of the fitting for both C$_{6}$D$_{6}$ and sTED detectors with $E_1$ and $E_2$ between 1 and 2 MeV. The region selected for the fitting procedure corresponds to the 4.9~eV $^{197}$Au($n$,$\gamma$) saturated resonance because of the large statistics and relatively steady count rate. The quality of the fit lends confidence on the extracted parameters $a_\circ$ and $\Delta t_{\circ}$.

\begin{figure}
    \centering
    \begin{tabular}{c}
    \includegraphics[width=\columnwidth]{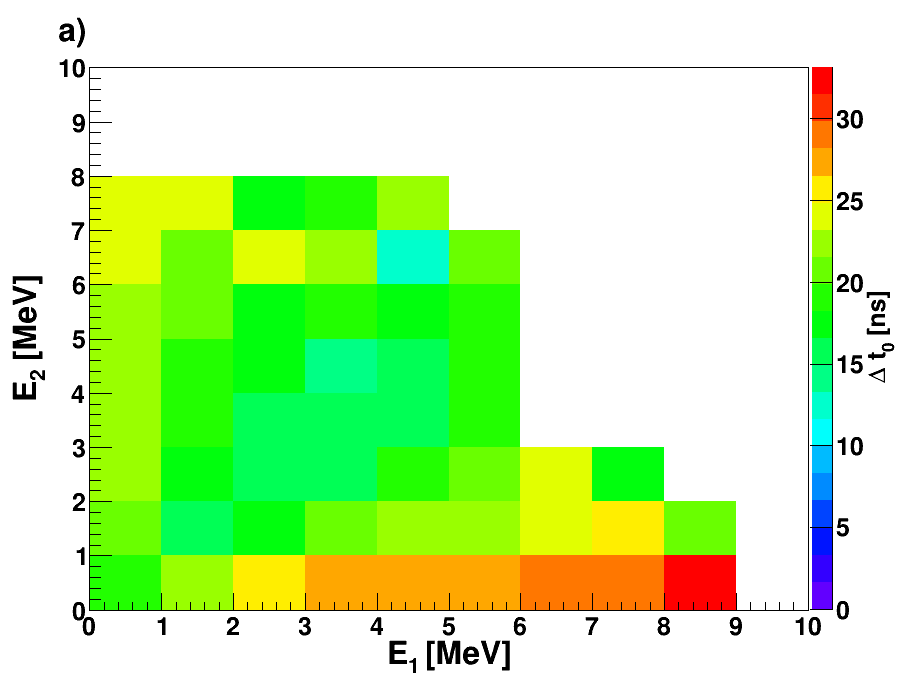} \\
    \includegraphics[width=\columnwidth]{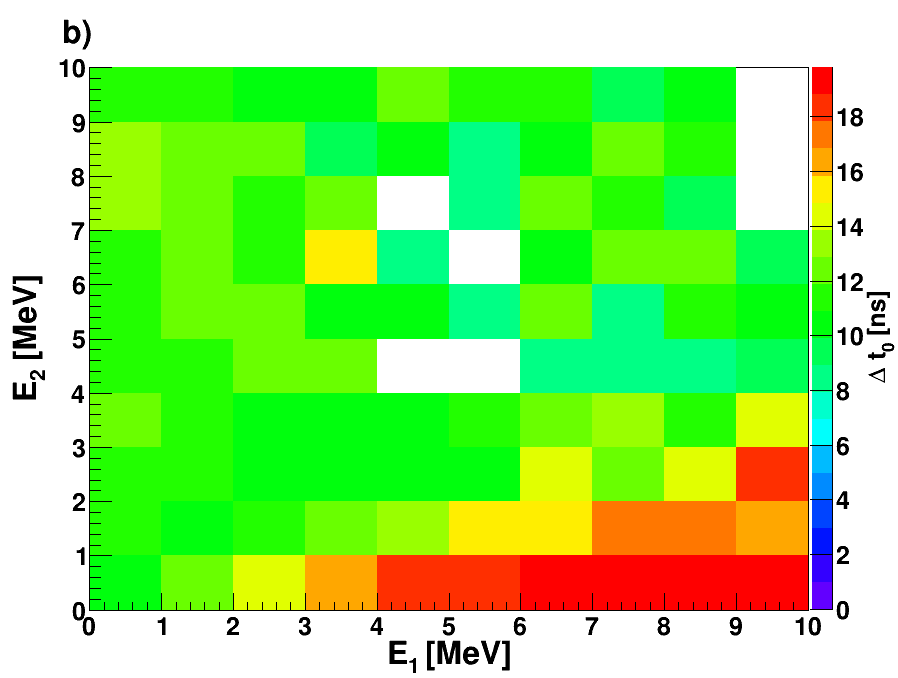} \\
    \end{tabular}
    \caption{$\Delta t_{\circ}(E_{1},E_{2})$ distribution calculated from the $\Delta t$ distribution fitting procedure. Panel a) C$_{6}$D$_{6}$ detector. Panel b) sTED module. White regions suffer from low statistics and could not be fitted.}
    \label{fig:Deltat0}
\end{figure}

\begin{figure}
    \centering
    \begin{tabular}{c}
    \includegraphics[width=\columnwidth]{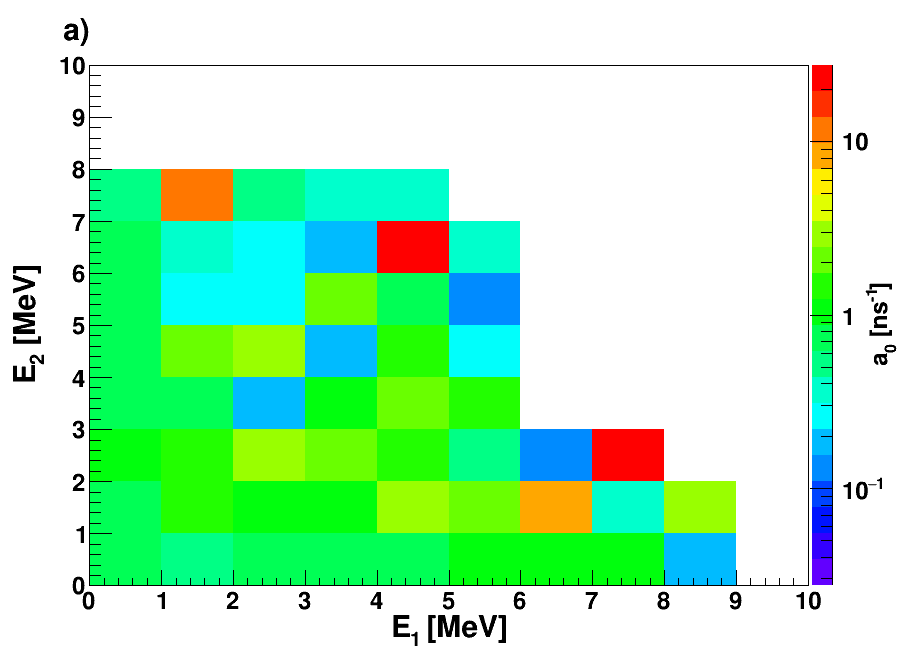} \\
    \includegraphics[width=\columnwidth]{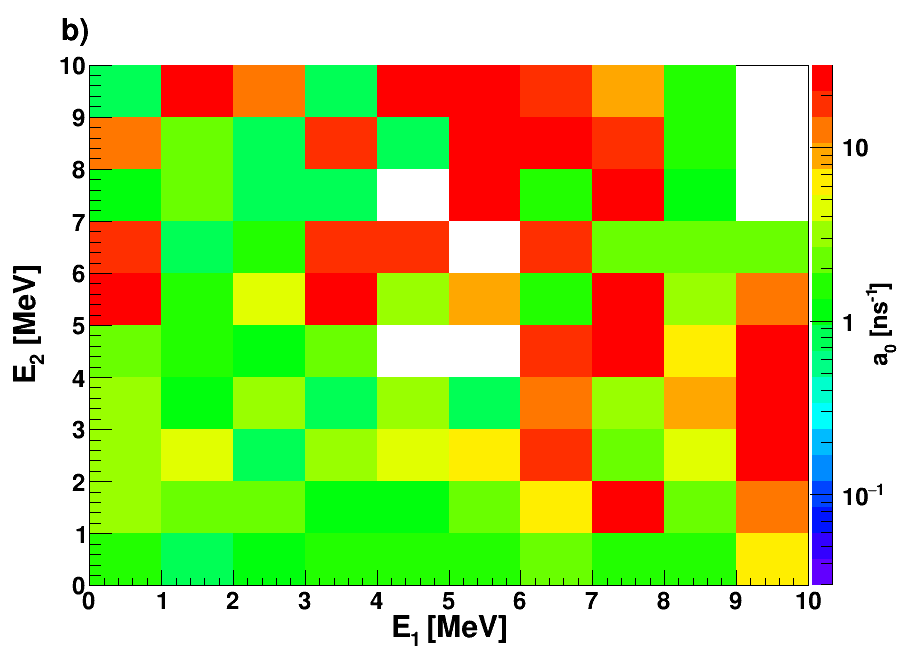} \\
    \end{tabular}
    \caption{$a_{0}(E_{1},E_{2})$ distribution calculated from the $\Delta t$ distribution fitting procedure for a) C$_{6}$D$_{6}$ and b) sTED module. White regions suffer from low statistics and could not be fitted.}
    \label{fig:a0}
\end{figure}

In practice, this fitting procedure needs to be applied to map $\Delta t_{0}(E_{1},E_{2})$ and $a(E_{1},E_{2})$ on a grid. However, limited statistics allows to use only (relatively) coarse bins. The results of the fits using 1~MeV wide intervals are given in Fig.~\ref{fig:Deltat0} and Fig.~\ref{fig:a0} for $\Delta t_{\circ}$ and $a_{\circ}$, respectively. For practical use, these distributions are fitted to a restricted bi-linear function of the form

\begin{equation}\label{eq:Delta}
    \Delta t_{\circ}(E_{1},E_{2})=\Delta t_{0}+b_{1}E_{1}+b_{2}E_{2},
\end{equation}

\begin{equation}\label{eq:a}
    a_{\circ}(E_{1},E_{2})=a_{0}+a_{1}E_{1}+a_{2}E_{2}.
\end{equation} 

Such fit smears fluctuations in individual bins and is thus better suited for practical use for any pair of $E_{1}$ and $E_{2}$ values. The bilinear fit was performed to signals ranging from 0 up to 7~MeV for $E_{1}$,$E_ {2}$, avoiding data-points of low statistics.

Fig.~\ref{fig:Deltat0} indicates that, in general, the smaller the ratio between $E_{2}$ and $E_{1}$, the smaller becomes also the detection probability of the second signal at given $\Delta t$. This feature is reflected in larger $\Delta t_{\circ}(E_{1},E_{2})$ values for smaller ratio, which is in agreement with previous works~\cite{MENDOZA2014,GUERRERO2015} and as indicated by Fig.~\ref{fig:Signal} it is not really surprising. In the case of C$_{6}$D$_{6}$ detectors, $\Delta t_{\circ}(E_{1},E_{2})$ ranges from about 12~ns up to a maximum value of 30~ns. For the smaller sTED modules this parameter spans from 8~ns up to 20~ns. Shorter $\Delta t_{\circ}$ for sTED may be ascribed to a better light-collection efficiency and faster PMT response of the smaller detector~\cite{electron_PMT,Hamamatsu_PMT}.

The $a_{\circ}(E_{1},E_{2})$ distribution for both detection systems, C$_{6}$D$_{6}$ and sTED module, is shown in Fig.~\ref{fig:a0}. On average, sTED shows significantly smaller $a_{0}$ values than the C$_6$D$_6$, with a detection probability recovery for the second signal, which is about 50\% faster when compared to the C$_{6}$D$_{6}$ detector. The parameters found for the bi-linear fit of Eq.(~\ref{eq:Delta}) and (~\ref{eq:a}) are reported in Tab.~\ref{tab:Delta0} and Tab.~\ref{tab:a0}, respectively.

\begin{table}[htb]
\centering
\begin{tabular}{c c c c}\hline
Detector & $\Delta t_{0}$ [ns] & $b_{1}$ [ns/MeV] & $b_{2}$ [ns/MeV] \\ \hline
sTED & 13.14 & 9.34$\times$10$^{-2}$ & -4.78$\times$10$^{-1}$ \\ 
C$_{6}$D$_{6}$ & 21.4 & 1.12$\times$10$^{-1}$ & -6.05$\times$10$^{-1}$ \\ \hline 
\end{tabular}
\caption{Parameters of the fit of Eq.~\ref{eq:a} shown in Fig.\ref{fig:Deltat0}.}
\label{tab:Delta0}
\end{table}
\begin{table}[htb]
\centering
\begin{tabular}{c c c c}\hline
Detector & $a_{0}$ [ns$^{-1}$] & $a_{1}$ [ns$^{-1}$/MeV] & $a_{2}$ [ns$^{-1}$/MeV] \\ \hline
sTED & 4.54$\times$10$^{-1}$ & 2.62$\times$10$^{-1}$ & 2.7$\times$10$^{-1}$ \\
C$_{6}$D$_{6}$ & 1.11 & -5.05$\times$10$^{-2}$ & -1.0$\times$10$^{-1}$ \\ \hline
\end{tabular}
\caption{Parameters of the fit of Eq.~\ref{eq:a} shown in Fig.\ref{fig:a0}.}
\label{tab:a0}
\end{table}

\subsubsection{Pile-up generator}\label{sec:pile-up_model}
Pile-up occurs for signals separated by a small time difference, generally shorter than the region where dead-time effects have a noticeable impact. Therefore, the pile-up effect is not  accessible by fitting the measured $\Delta t$ distribution of two consecutive pulses and an alternative approach needs to be implemented. To this aim, we assume that if the time difference between two consecutive signals is smaller than a constant value $\Delta t_{p}$, then the energy of the first signal is considered to be the sum of both pile-up signal energies and the contribution of the second signal is removed from the event list. 

In order to determine a reasonable value for $\Delta t_{p}$, the procedure described at the beginning of Sec.~\ref{sec:Model} was tested for several value. We have adopted the one that gives the most accurate reproduction of the experimental deposited-energy spectra for different count rates. These different count rates are reached by separate treatment of low- and high-intensity proton pulses. The best $\Delta t_{p}$ values obtained are 18~ns and 8~ns for C$_{6}$D$_{6}$ and sTED module, respectively. These quantities are comparable or smaller than the pulse-width of both C$_{6}$D$_{6}$ and sTED module. The comparison of simulated $h_d(E)$ spectrum with the best $\Delta t_p$ to the experimental one in the reference case of 4.9 eV $^{197}$Au($n$,$\gamma$) is shown in Fig.~\ref{fig:Deposited_spectra}. The fact that both count rate deposited energy spectra become now compatible regardless of the proton-intensity or count rate level, lend confidence on the applied methodology.

\subsection{Dead-time and pile-up correction methodology for constant count rate $r$}\label{sec:correction}

Dead-time corrections estimate count losses above a certain deposited energy detection threshold $E_{thr}$ at given count rate $r$~\cite{Knoll1979}. This is equivalent to requiring that the integral of undistorted deposited spectra is equal to the integral of the distorted deposited energy spectra times dead-time correction $f_{dt}$,

\begin{equation}\label{eq:CountsIntegral}
    \int_{E_{thr}}{h_{MC}(E)dE}=f_{dt}(E_{thr},r) \int_{E_{thr}}{h_{d}(E)dE}.
\end{equation}
Thus, the dead-time correction is deduced from the ratio of both integrals

\begin{equation}\label{eq:Correction}
    f_{dt}(E_{thr},r)=\frac{\int_{E_{thr}}{h_{MC}(E)dE}}{\int_{E_{thr}}{h_{d}(E)dE}}.
\end{equation}

The actual determination of the cross section using the TED principle requires a detection efficiency proportional to the deposited energy. The latter is obtained by means of the application of a weighting function $W_{f}(E)$ to the spectrum~\cite{MacKlin1967,Abbondanno2004,Borella07}, i.e. application of the PHWT assuming no pile-up. The number of detected neutron capture reactions for a particular isotope ($n_{R}$) is equal to the integral of deposited energy spectra weighted by $W_{f}(E)$ times a normalization factor $C$. In absence of dead-time and pile-up, the deposited energy spectrum corresponds to $h_{MC}(E)$. Thus, 
\begin{equation}
n_{R}=C\int_{E_{thr}}{h_{MC}(E)W_{f}(E)dE}
\end{equation}

In practice, the measured energy spectrum, $h_{d}(E)$, is not only affected by dead-time effects but also by pile-up events. The actual number of detected reactions ($n^{\prime}_{R}$) is then given by

\begin{equation}
n^{\prime}_{R}=C\int_{E_{thr}}{h_{d}(E)W_{f}(E)dE}
\end{equation}

Experimentally, one deals with non-ideal detectors that under certain count-rate conditions are affected by dead-time and pile-up effects and thus $h_{MC}(E)\neq h_{d}(E)$ and $n_{R}\neq n^{\prime}_{R}$. In other words, the detection efficiency to the particular ($n$,$\gamma$) is not constant anymore, even if the TED principle holds, because the deposited energy spectra has changed. In analogy with $f_{dt}(E_{thr},r)$, we can introduce a factor $f_{dtpu}(E_{thr},r)$ that describes, for a constant count rate $r$, the size of the correction that needs to be applied in order to recover the detection efficiency to the undistorted value. Therefore, for a constant count rate, the requirement $n_{r}=n^{\prime}_{r}$ transforms to
\begin{equation}\label{eq:Q-Correction}
\int_{E_{thr}}{h_{MC}(E)W_{f}(E)dE} = f_{dtpu}(E_{thr},r) \int_{E_{thr}}{h_{d}(E)W_{f}(E)dE}.
\end{equation}
And as for dead-time correction, the proposed correction can be quantified from the ratio of both integrals,

\begin{equation}\label{eq:Q-Correction}
    f_{dtpu}(E_{thr},r)=\frac{\int_{E_{thr}}{h_{MC}(E)W_{f}(E)dE}}{\int_{E_{thr}}{h_{d}(E)W_{f}(E)dE}}.
\end{equation}

In reality, $r$ is not experimentally accessible because of the detection threshold. Nevertheless, we can use measured count rate $r^{\prime}$, the number of counts detected above the detection threshold per time unit as for a given cascade there is a unique relation between $r$ and $r^{\prime}$ and the correction factors can be equivalently formulated using $r^{\prime}$ instead of $r$.

\begin{figure}
    \centering
    \includegraphics[width=\columnwidth]{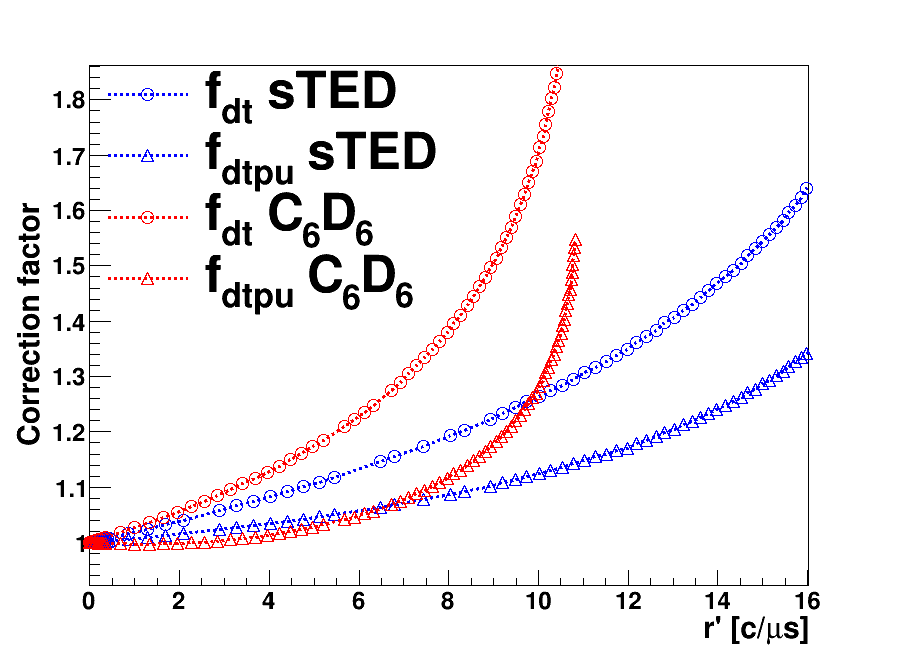} \\
    \caption{$f_{dt}$ and $f_{dtpu}$ corrections for both, C$_{6}$D$_{6}$ and sTED module, using a detection threshold of 300~keV and 200~keV, respectively.}
    \label{fig:Correction}
\end{figure}

Fig.~\ref{fig:Correction} shows both $f_{dt}$ and $f_{dtpu}$ corrections for $^{197}$Au($n$,$\gamma$) as a function of the measured count rate $r^{\prime}$ for both sTED and C$_{6}$D$_{6}$ detectors. Thresholds for sTED and C$_{6}$D$_{6}$ detectors are set to 200~keV and 300~keV, respectively. Both corrections are negligible (below 0.5\%) only for $r^{\prime}\lesssim$0.1~c/$\mu s$. Further $f_{dt}$ and $f_{dtpu}$ corrections are almost, with values in the order of 1-2\%, identical at very low values of $r^{\prime}\lesssim$0.5 c/$\mu s$. For these low count rates the pile-up probability is still low and thus the fraction of missing counts is small. As $r^{\prime}$ increases, the pile-up starts to introduce deviations in the deposited energy spectrum, with respect to no pile-up case. In practice, the correction on the pile-up partially compensates for the dead-time one due to the properties $W_{f}(E)$ that gives larger weight to pulses with higher $E$. Considering both these effects is thus critical when aiming at high precision measurements with TED detectors.

The range of applicability of the proposed method depends on the specific detector pulse-shape characteristics. The numerical limitation of the model for C$_{6}$D$_{6}$ and sTED are $\sim$11 c/$\mu$s and 16~c/$\mu s$, respectively. Based on analysis of $^{197}$Au($n$,$\gamma$) experimental data, see Sec.~\ref{sec:results}, we can make a conclusion that the method is applicable up to r$^{\prime}\approx$~8 c/$\mu s$, which corresponds to $f_{dt}\approx$~1.35 for C$_{6}$D$_{6}$ detector. For sTED module we checked the maximum $r^{\prime}\approx$~8.2 c/$\mu s$ without spotting any problems in the applicability of the approach. In analogy with C$_{6}$D$_{6}$, we believe that the method should work again at least in the range $f_{dt} \lesssim 1.35/1.4$, i.e. up to at least $\approx$ 12-14~c/$\mu s$ for this detector. These limits should be avoided because small uncertainties in the corrections can be translated in large systematic uncertainties to the actual deduced yield. For this reason we propose a use of the current dead-time and pile-up model for C$_{6}$D$_{6}$ and sTED up to count rates about $\sim$6~c/$\mu s$ and $\sim$10~c/$\mu s$, respectively.

It is worth mentioning that although the parameters, $a_{\circ}$, $\Delta t_{\circ}$ and $\Delta t_{p}$, of the dead-time and pile-up models have been adjusted using the 4.9 eV saturated resonance in $^{197}$Au($n$,$\gamma$), they do not depend on the specific $\gamma$-ray cascade details or on the count rate conditions and the main dependency on these parameters arises from the detector response and pulse-shape characteristics. On the other hand, since the actual $h_{MC}(E)$ and $h_{d}(E)$ spectra depend on the actual isotope under study, the particular dead-time and pile-up corrections will be the isotope specific too. However, they can be quantified by simple repeating the MC procedure described in the beginning of Sec.~\ref{sec:Model} adopting $a_{\circ}$, $\Delta t_{\circ}$ and $\Delta t_{p}$ parameters derived in this paper.

\subsection{Correction for rapidly varying count rates}\label{sec:corr_variation}

For any ($n$,$\gamma$) cross section measurement in the resolved resonance region using ToF, $r^{\prime}$ varies strongly as a function of the neutron energy (E$_{n}$) due to different strengths and positions of the neutron resonances. Under such rapidly changing $r^{\prime}$ conditions, $f_{dt}$ and $f_{dtpu}$ corrections (Sec.~\ref{sec:correction}) can be applied independently to individual neutron energy bins only assuming a constant $r^{\prime}$ rate within the bin. Usually, this condition is achieved by using a neutron energy binning sufficiently narrow compare to the corresponding $r^{\prime}$ variation rate~\cite{GUERRERO2015}. In this way, a constant $r^{\prime}$ can be considered and dead-time and pile-up corrections can be applied to individual bins using the methodology explained in Sec.~\ref{sec:correction}.
\section{Validation of dead-time and pile-up correction}\label{sec:results}

The methodology described in Sec.~\ref{sec:correction} and~\ref{sec:corr_variation} has been validated using $^{197}$Au($n$,$\gamma$) experimental data from an experiment performed at n\_TOF EAR2 utilizing C$_{6}$D$_{6}$ detector and sTED modules. The Au target, with a size of 20~mm diameter and 0.1~mm thickness, was placed in the center of the setup. sTED and C$_{6}$D$_{6}$ detectors were placed at a distance from the target of 4.5~cm and 17~cm, respectively. In this way the count rates detected by both detection systems are comparable. For further details about the experiment, the reader is referred to Refs.~\cite{Lerendegui2023,Balibrea22-NPA}. This specific reaction has been chosen because of its importance for ($n$,$\gamma$) cross section normalization. The strong 4.9~eV resonance in $^{197}$Au($n$,$\gamma$) is commonly used in many experiments for normalization~\cite{Abbondanno2004,MacKlin1969,Massimi2011}. During the experiment, both LI and HI pulses were delivered to the spallation target, yielding two different $r^{\prime}$ situations, that can be compared at each neutron-energy bin. In this work, reaction yield is defined as the number of ($n$,$\gamma$) reactions detected as a function of the neutron energy. Because this quantity should not depend on the proton intensity when properly normalized, after applying the corresponding dead-time and pile-up corrections, both corrected reaction yields should coincide within the experimental uncertainty at each energy bin.

\begin{figure*}
    \centering
    \begin{tabular}{c c}
    \includegraphics[width=0.67\columnwidth]{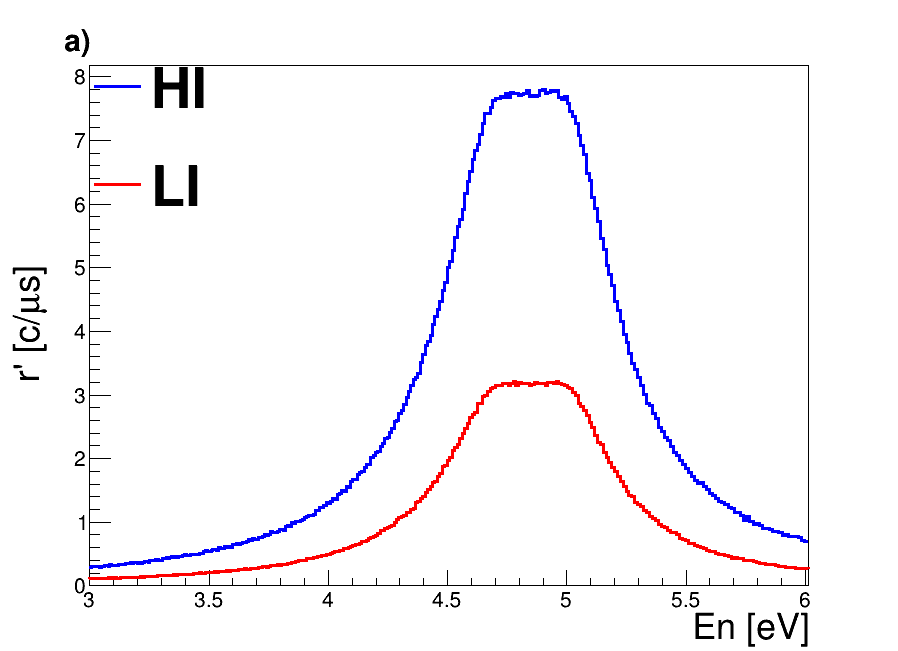} &
    \includegraphics[width=0.67\columnwidth]{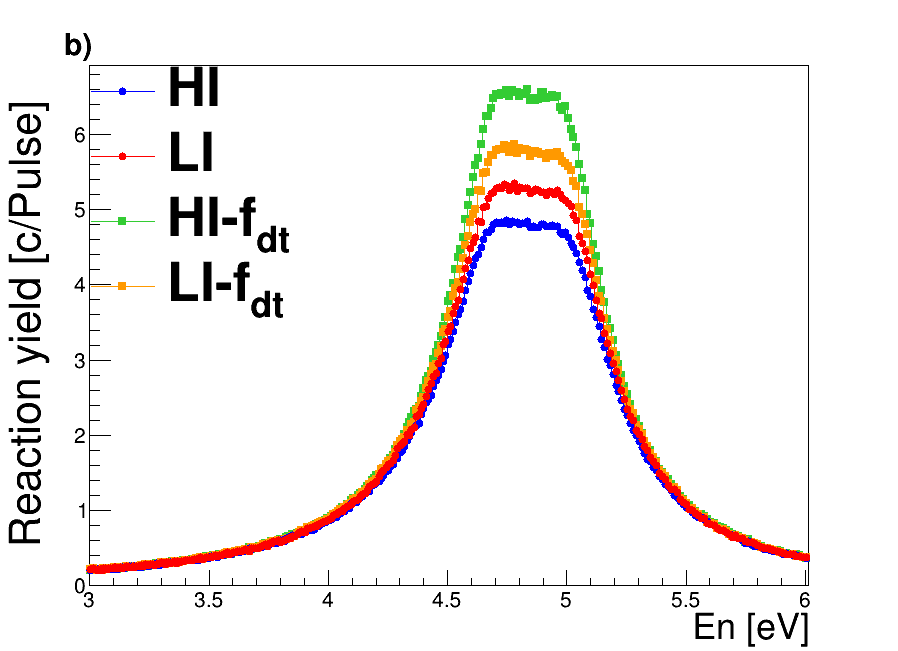} \\
    \includegraphics[width=0.67\columnwidth]{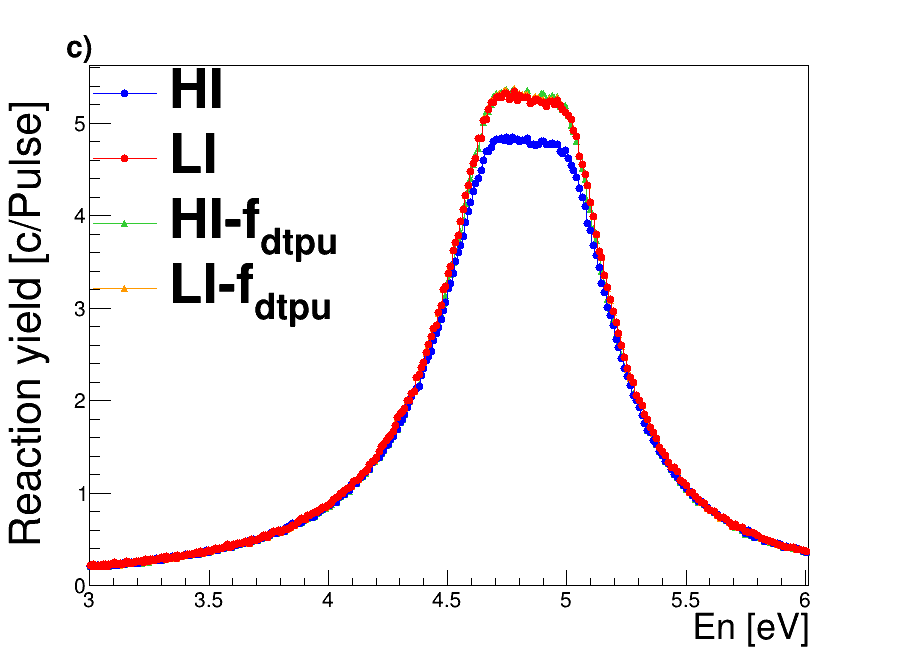} &
     \includegraphics[width=0.67\columnwidth]{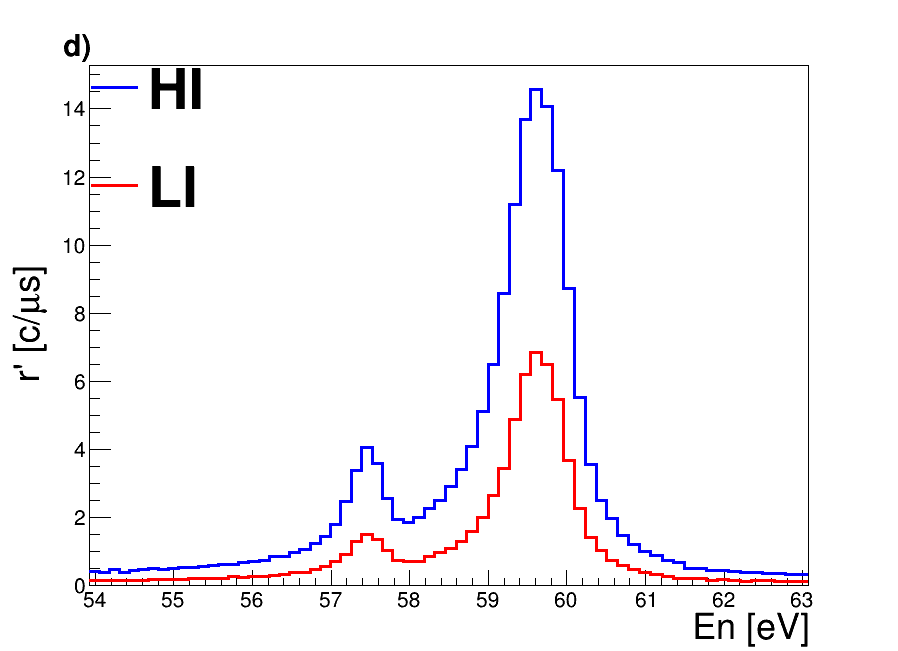} \\
    \includegraphics[width=0.67\columnwidth]{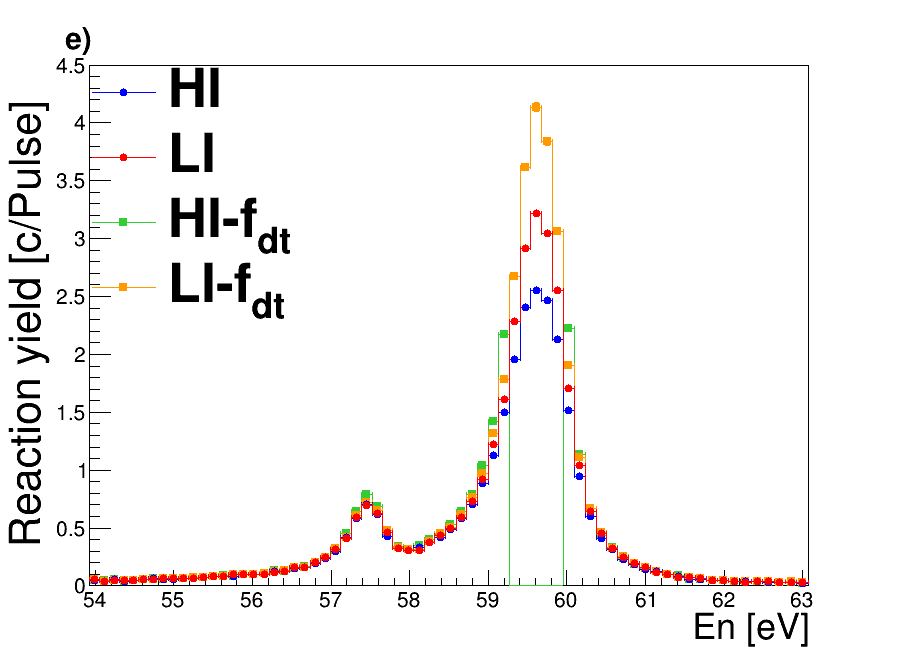} &
    \includegraphics[width=0.67\columnwidth]{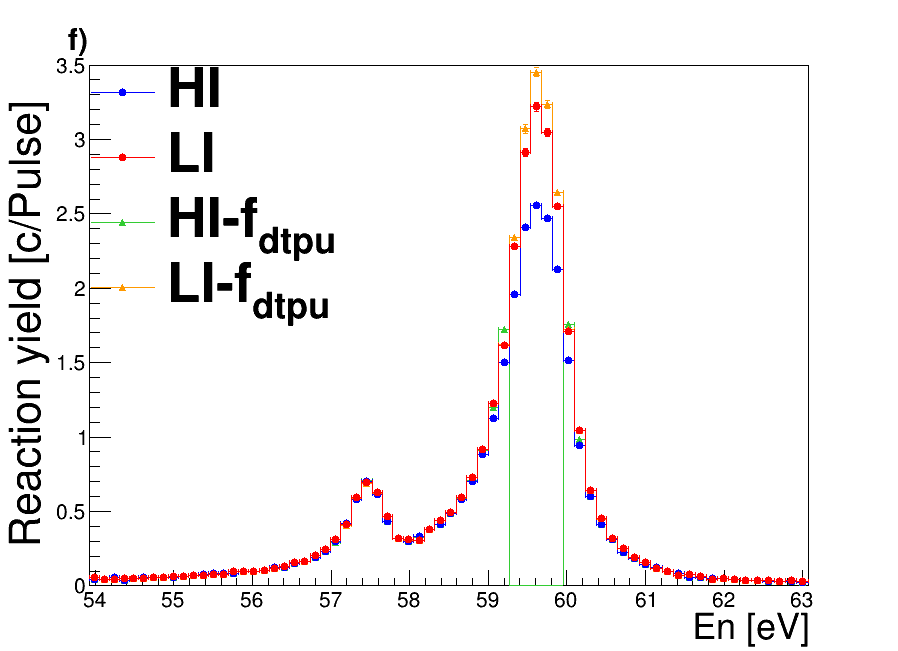} \\
    \end{tabular}
    \caption{Performance of $f_{dt}$ and $f_{dtpu}$ corrections applied to C$_{6}$D$_{6}$ in the neutron energy range from 3 to 6~eV and 54 to 63~eV. Panels a) and d) present $r^{\prime}$ as a function of neutron energy for HI and LI pulses. Panels b) and e) give corresponding reaction yield from experiment and after application of $f_{dt}$ correction, while panels c) and f) after application of $f_{dtpu}$ correction.}
    \label{fig:ToF_S_resonance_C6D6}
\end{figure*}

Two different neutron-energy regions were selected for illustration purposes: the neutron energy range from 3~eV to 6~eV where the saturated $^{197}$Au($n$,$\gamma$) resonance at 4.9 eV is located, and the region from 54 eV to 63~eV corresponding to the largest peak count rate resonance in $^{197}$Au($n$,$\gamma$) at 60~eV.

\begin{figure*}
    \centering
    \begin{tabular}{c c}
    \includegraphics[width=0.67\columnwidth]{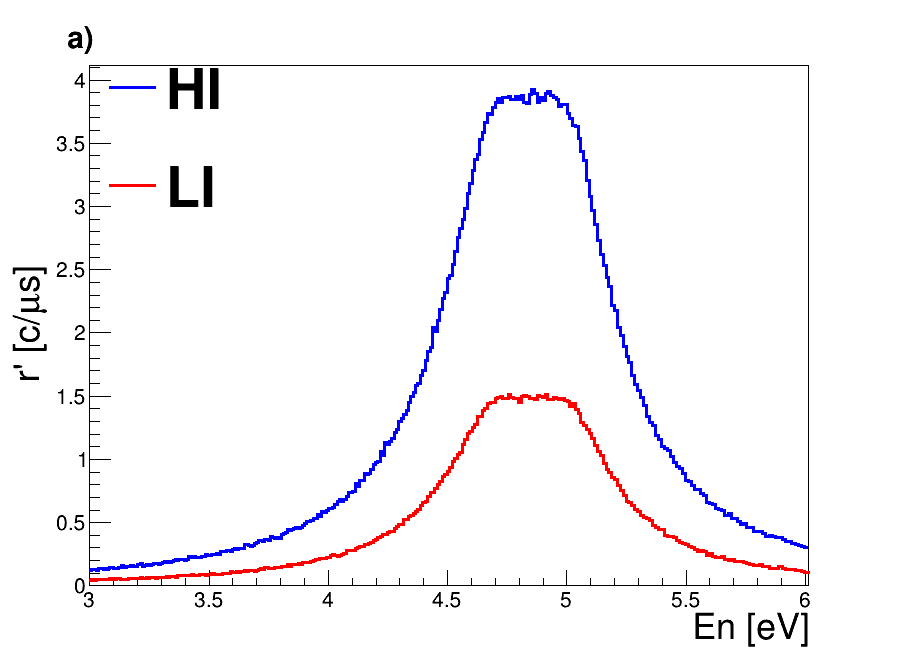} &
    \includegraphics[width=0.67\columnwidth]{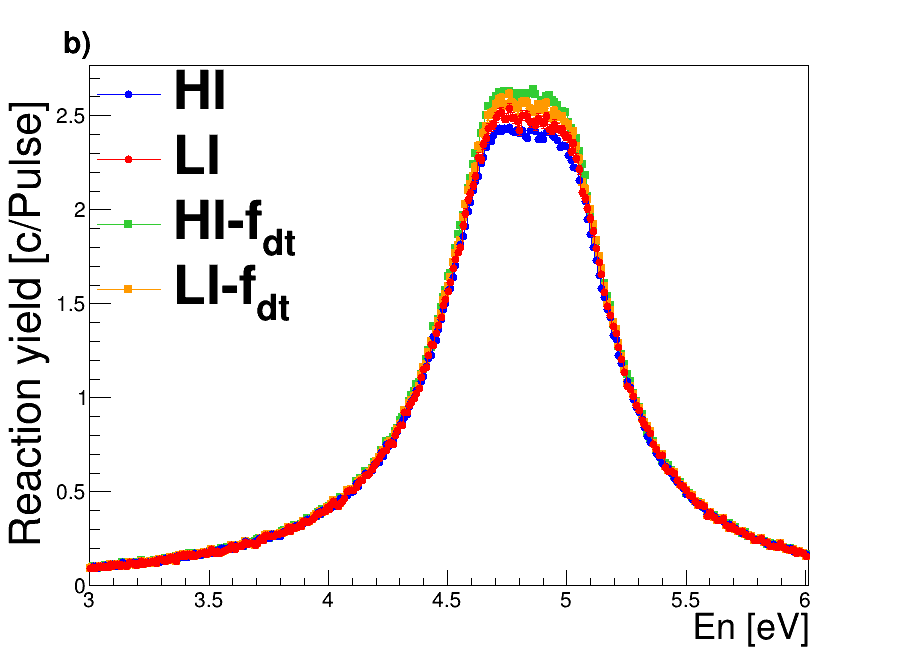} \\
    \includegraphics[width=0.67\columnwidth]{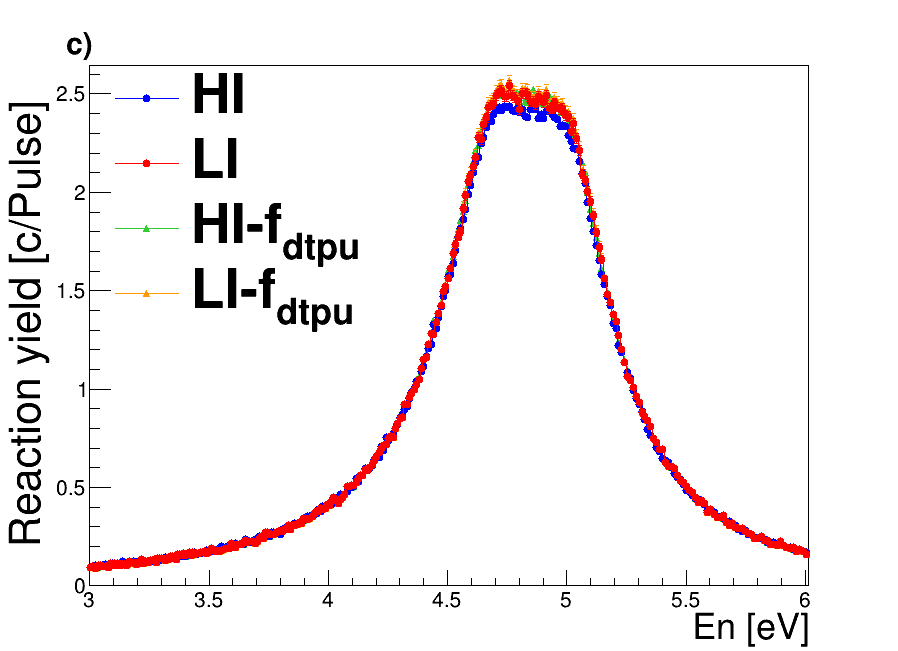} &
    \includegraphics[width=0.67\columnwidth]{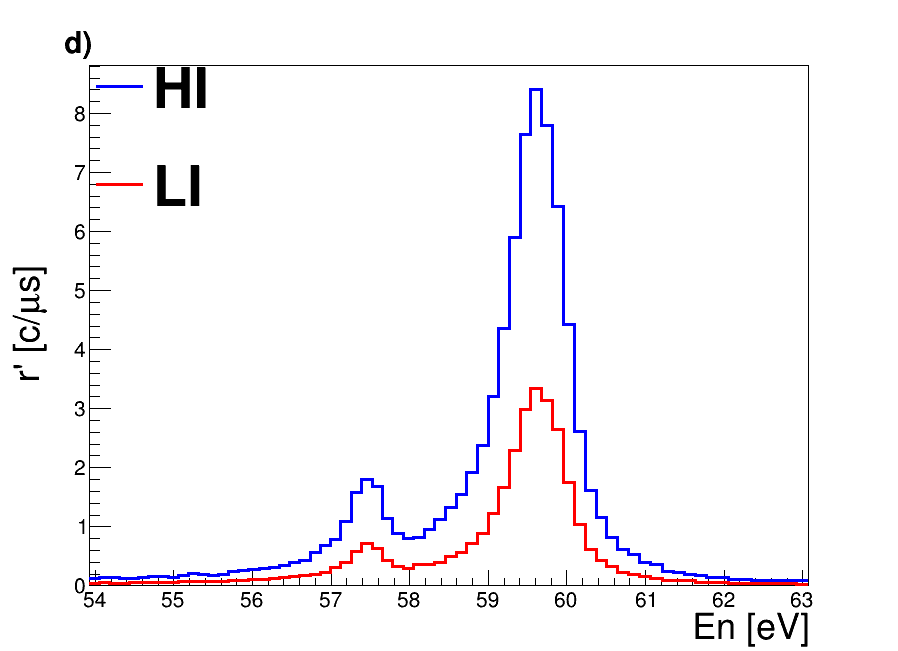} \\
    \includegraphics[width=0.67\columnwidth]{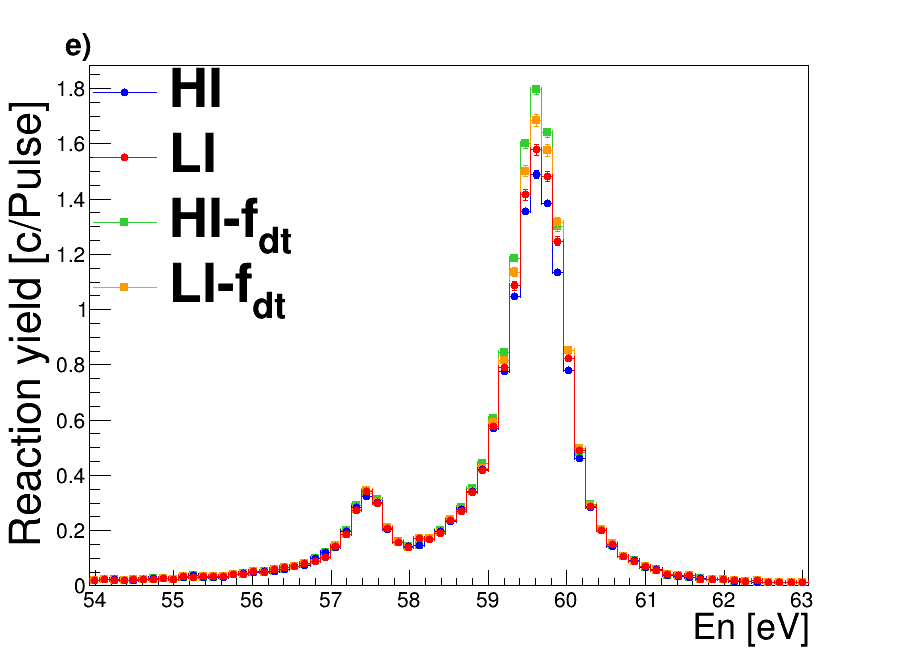} &
    \includegraphics[width=0.67\columnwidth]{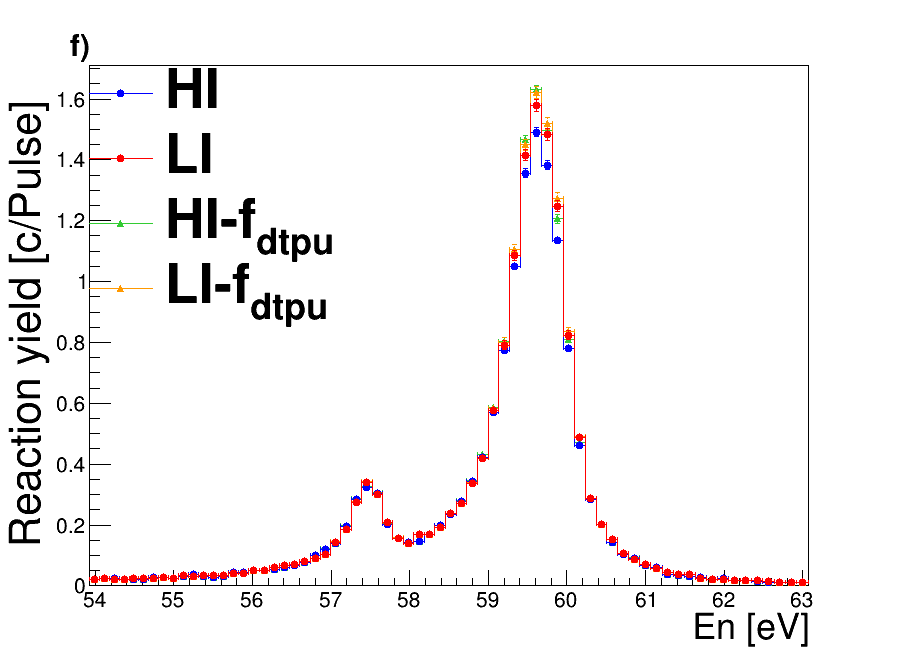} \\
    \end{tabular}
    \caption{Performance of $f_{dt}$ and $f_{dtpu}$ corrections applied to sTED in the neutron energy range from 3 to 6~eV and 54 to 63~eV. Panels a) and d) present $r^{\prime}$ as a function of neutron energy for HI and LI pulses. Panels b) and e) give corresponding reaction yield from experiment and after application of $f_{dt}$ correction, while panels c) and f) after application of $f_{dtpu}$ correction.}
    \label{fig:ToF_S_resonance_sTED}
\end{figure*}

The results, as a function of neutron energy, are shown in Fig.~\ref{fig:ToF_S_resonance_C6D6} for C$_6$D$_6$ and in Fig.~\ref{fig:ToF_S_resonance_sTED} for sTED, respectively. In both figures, panels a) and d) display $r^{\prime}$ in the two selected neutron energy ranges, respectively. Panels b) and e) show the reaction yields together with the $f_{dt}$-corrected one. Finally, panels c) and f) show the same measured yields and $f_{dtpu}$-corrected ones.

The HI and LI C$_6$D$_6$ uncorrected experimental reaction yield at the flat top part of the saturated 4.9~eV resonance reveals differences as large as $\sim$8\%, as shown in panels b) and c) of Fig.~\ref{fig:ToF_S_resonance_C6D6}. The differences observed are indeed caused by by different contribution contribution of dead-time and pile-up effects during HI and LI pulses. Based on Fig.~\ref{fig:Correction} for the count rates $r^{\prime}$, 3~c/$\mu$s and 7.5~c/$\mu$s, $f_{dt}$ corrections have values close to 1.1 and 1.3, respectively. After application of dead-time correction $f_{dt}$ (panel b), differences between differences between HI and LI corrected yields actually increase up to $\sim$9\%. 

The situation changes when $f_{dtpu}$ corrections are applied. This is demonstrated in panel c), where both HI and LI corrected yields show a satisfactory agreement. These yields are close to the uncorrected yield obtained with LI pulses.This is evident from Fig.~\ref{fig:Correction} as the $f_{dtpu}\sim$1.01 for $r^{\prime}\sim$3 c/$\mu s$. One can therefore conclude from Fig.~\ref{fig:ToF_S_resonance_C6D6}, that the $f_{dtpu}$ correction is able to properly account for deviations observed in the deposited energy spectra arising from severe pile-up and dead-time effects present in the saturated 4.9 eV resonance up to count rate of (at least) about 7.5~c/$\mu$s.

Fig.~\ref{fig:ToF_S_resonance_C6D6}-b), d), f) shows the other selected neutron energy range, which actually corresponds to even more stressful count rate situation. Measured count rates of $r^{\prime}\sim$7~c/$\mu$s and 15.0~c/$\mu$s are reached for LI and HI pulses, respectively, at the top of the 60 eV resonance. In fact, $r^{\prime}\sim$15.0~c/$\mu$s can be regarded as beyond the correction capabilities of the current algorithm, as discussed in Sec.~\ref{sec:correction} and shown in Fig.~\ref{fig:Correction}.

Such large count rate situations would require a more sophisticated model. Another option is a use of detector system which allows reduction of the count rate while preserving similar statistics, a solution like an array of sTED modules, or a system with the response that reduces dead-time and pile-up effects. The current limitation for C$_{6}$D$_{6}$ detectors is clearly observed near the top of the 60~eV resonance, where both $f_{dt}$ and $f_{dtpu}$ corrections fail to provide a satisfactory results for HI pulses. sTED results for the two neutron-energy ranges are then shown in Fig.~\ref{fig:ToF_S_resonance_sTED}. The smaller active volume of sTED in comparison to C$_{6}$D$_{6}$ reduces significantly the count rates. As shown in panel a), in the saturated 4.9~eV resonance of $^{197}$Au($n$,$\gamma$) the maximum $r^{\prime}$ during LI and HI proton pulses are $\approx$ 1.4~c/$\mu$s and 3.9~c/$\mu$s, respectively. In this case, the differences observed between LI and HI registered yields is about 5\%. As shown in Fig.~\ref{fig:Correction}, $f_{dt}$ corrections for LI and HI pulses are 5\% and 15\%, respectively. The $f_{dtpu}$ values amounts to 3\% and 1\%, respectively. Again, if only the $f_{dt}$ correction is applied (see panel b), the corrected yields still differ by about 2\%. Only when the full $f_{dtpu}$ correction is applied (see panel c), deviations become only 0.5\%, which is within the statistical uncertainty of the measured yield.

In the neutron energy range between 54 and 63~eV the maximum $r^{\prime}$ measured with sTED are of 3.0~c/$\mu$s and 8.2~c/$\mu$s for LI and HI pulses, respectively (see Fig.~\ref{fig:ToF_S_resonance_sTED}-d) and Fig.~\ref{fig:ToF_S_resonance_sTED}-f)). The model works nicely even for these large count rates, obtaining results with deviations within errors on the corrected yield from both LI and HI pulses. The wider range of applicability can be ascribed to the narrower signals of sTED when compared to C$_{6}$D$_{6}$ detectors. Again, only the $f_{dt}$ correction is not sufficient for the 60~eV resonance as shown in panel e). 

Finally, it is worth emphasizing that it is preferable to design neutron-capture experiments in such a way, that corrections as large as those discussed here are avoided. However, in situations where this is not possible -- as those typically performed in the n\_TOF EAR2 -- the presented method will represent a suitable solution to account for high count rate conditions. 

\section{Summary and conclusions}\label{sec:Summary_and_Conclusions}

Dead-time and pile-up effects represent a conspicuous source of systematic error in neutron-capture experiments dealing with very high count rates. The interplay between these two effects and their impact to the final capture cross-section requires of a sophisticated correction algorithm. Dead-time effects lead to an underestimation of the reaction yield whereas pile-up can induce an apparent increase in detection efficiency that, if not properly treated, would lead to biased ($n$,$\gamma$) cross section result after the PHWT is applied.

In this work we have proposed and validated experimentally a Monte Carlo method for detectors using Total Energy Deposition principle (which is obtained via application of PHWT) that considers dead-time and pile-up effects simultaneously in a consistent way. We have applied it to two different types of C$_{6}$D$_{6}$ liquid scintillators at different count rate conditions and demonstrated the capability to account for these effects in neutron-capture cross-section measurements under very high count rate conditions that can easily be reached at n\_TOF EAR2. The parameters needed for the implementation of dead-time correction ($f_{dt}$) can be obtained experimentally and are independent of $\gamma$-rays emitted during individual decays following the neutron capture. On the other hand, the correction that takes into account also pile-up ($f_{dtpu}$) depends on the $\gamma$ decay pattern when the PHWT is used as the the deduced energy of events coming from the pile-up depends on this sequence. In any case, we showed that both the corrections are important. For this reason, the process described in Sec.~\ref{sec:Model} will have to be repeated for the particular reaction under study.

The proposed correction method was tested on data from $^{197}$Au($n$,$\gamma$) data using two different detectors, a conventional, large volume, carbon fiber  C$_{6}$D$_{6}$ with a volume of 1~L and much smaller sTED module with a volume of 49~mL. We found that the method brings very good results if it is applied to experimental count rates up to $\sim$8~c/$\mu s$ for both detection systems with differences of 1-2\% between corrected HI and LI proton pulses. This coun trate is close to the upper limit of applicability for large-volume C$_{6}$D$_{6}$ as we see that for more than about 10-11 c/$\mu$s the method does not fully work. The range of applicability for sTED is larger attending to the results shown in Fig.~\ref{fig:Correction}. However, higher count rates, even after applying dead-time and pile-up corrections can lead to significant systematic errors that, after propagated, can lead to biased ($n$,$\gamma$) cross section values. The different range of applicability of the presented model is a result of the different pulse-shape characteristics of different detection systems. It is worth to recall that this type of corrections may be crucial when applying the saturated-resonance technique for neutron capture yield normalization~\cite{MacKlin1967,Abbondanno2004,Borella07,SCHILLEBEECKX2012} if a systematic uncertainty on the level of at most a few percent is targeted in the determination of the neutron-capture cross section.

\section*{Acknowledgment}
This work has been carried out in the framework of a project funded by the European Research Council (ERC) under the European Union's Horizon 2020 research and innovation programme (ERC Consolidator Grant project HYMNS, with grant agreement No.~681740). This work was supported by grant ICJ220-045122-I funded by MCIN/AEI/10.13039/501100011033 and by European Union NextGenerationEU/PRTR. The authors acknowledge support from the Spanish Ministerio de Ciencia e Innovaci\'on under grants PID2019-104714GB-C21, PID2022- and the funding agencies of the participating institutes. The authors acknowledge the financial support from MCIN and the European Union NextGenerationEU and Generalitat Valenciana in the call PRTR PC I+D+i ASFAE/2022/027

\bibliographystyle{elsarticle-num} 
\bibliography{bibiliography}





\end{document}